\documentclass[twocolumn, superscriptaddress, preview, amsmath, amssymb, aps, letterdraft, notitlepage,longbibliography]{revtex4-1}
\usepackage[utf8]{inputenc}
\usepackage{natbib}
\usepackage{graphicx}
\usepackage{mathtools}
\usepackage{xcolor}

\newcommand{\iu}{\mathrm{i}} 
\newcommand{\bra}[1]{\langle#1|}
\newcommand{\ket}[1]{|#1\rangle}
\def\>{\rangle}
\def\<{\langle}
\newcommand{\bk}[2]{\<{#1}|{#2}\>}

\begin{document}
\title{A Neural-Network Variational Quantum Algorithm for Many-Body Dynamics}

\author{Chee Kong Lee}
\email{cheekonglee@tencent.com}
\affiliation{Tencent America, Palo Alto, CA 94306, United States}
\author{Pranay Patil}
\affiliation{Laboratoire de Physique Th\'eorique, IRSAMC, Universit\'e de Toulouse, CNRS, UPS, France}
\author{Shengyu Zhang} 
\affiliation{Tencent Quantum Lab, Shenzhen, Guangdong 518057, China}
\author{Chang Yu Hsieh}
\affiliation{Tencent Quantum Lab, Shenzhen, Guangdong 518057, China}

\begin{abstract}
We propose a neural-network variational quantum algorithm to simulate the time evolution of quantum many-body systems.
Based on a modified restricted Boltzmann machine (RBM) wavefunction ansatz, the proposed algorithm can be efficiently implemented in near-term quantum computers with low measurement cost. Using a qubit recycling strategy, only one ancilla qubit is required to represent all the hidden spins in an RBM architecture.
The variational algorithm is extended to open quantum systems by employing a stochastic Schr\"{o}dinger equation approach.
Numerical simulations of spin-lattice models demonstrate that our algorithm is capable of capturing the dynamics of closed and open quantum many-body systems with high accuracy without suffering from the vanishing gradient (or `barren plateau') issue for the
considered system sizes.
\end{abstract}

\maketitle

\section{Introduction}
Accurate and efficient simulation of quantum many-body dynamics remains one of the most challenging problems in physics, despite nearly a century of progress. 
Renewed interest has been sparked in this field due to recent experiments with
Rydberg atoms\cite{Ryd,Ryd2}, which suggest the existence of scar states which do not thermalize.
This has lead to new studies of fragmented Hilbert spaces for such constrained models\cite{frag,frag2,frag3},
along with further studies on fractons, which are restricted excitations which can disperse only in
certain directions\cite{FracRev,Frac2}. These studies also tie in to the more established field of many
body localization\cite{MBLAlet,MBL2,MBL3}, which studies the possibility of extremely slow relaxation of high
energy states in systems with strong disorder. As many of the above phenomena are hard to study
analytically, there is a strong motivation to develop powerful numerical tools to further our
understanding.

One of the most powerful numerical tools at the disposal of condensed matter theorists is quantum
Monte Carlo, which has performed remarkably well for equilibrium physics of numerous
systems\cite{suzuki1993quantum,QMC2}.
This has made important the applicability of this technique to study real time dynamics. This
is often impossible due to the infamous sign problem\cite{Marshall,sign2}, and one of the few promising
ways in which practitioners have attempted to avoid this is by transferring the real time behavior
to functions which form coefficients in the wavefunction. These functions then need to have a
variational form which can be optimised to get reasonably good results on small
systems\cite{carleo2017unitary,ido2015time}. Even though one can get around the sign problem for these cases, severe ergodicity
restrictions in the Monte Carlo updates may render them inefficient, and necessitate specialized 
algorithms\cite{biswas2016quantum,sandvik,QDalgo}.
To allow variational wavefunctions a higher degree of expressibility, some ideas from machine 
learning, such as restricted Boltzmann machines (RBM), have been used\cite{Carleo2017a, nagy2019variational, glasser2018neural, Schmitt2019, Gutierrez2019, deng2017machine, sarma2019machine} to serve as a representation.
This has lead to a well-controlled way of approximating complicated wavefunctions with rich spatial
features. Neural networks have also been used to simulate open quantum
systems, which are numerically more challenging to study than closed systems, and promising
results have been achieved for both
dynamical\cite{Hartmann2019} and steady state\cite{nagy2019variational,vicentini2019variational, yoshioka2019constructing} features.

Due to recent advances in quantum computing, it has become possible to program a small number of
qubits to directly represent a quantum system using Noisy Intermediate-Scale Quantum (NISQ)
technology\cite{preskill2018quantum, Arute2019}. One of the many applications of this set up is to
speed up the optimization step for variational wavefunctions\cite{mcclean2016theory, Peruzzo2014, Farhi2014, kandala2017hardware, hempel2018quantum, Colless2018}. This serves as a substantial
improvement for cases where variational Monte Carlo is inefficient. Direct variational optimization
of the time-dependent Schr\"{o}dinger equation\cite{Li2017, Yuan2019, Heya2019, Lee2021} has also shown promise, and a large number of general
processes can be mapped on to this technique\cite{Endo2020}.

In this work, we engineer a neural-network variational quantum algorithm to simulate the dynamics of quantum many-body systems.
The algorithm integrates the power of an
RBM representation of quantum states with a quantum speed-up
coming from transferring the computationally heavy step of calculating expectation values on to the
quantum computer. 
We show that the variational algorithm can be extended to the dynamics of open quantum systems using a stochastic Schr\"{o}dinger equation approach.
The proposed method is benchmarked against canonical spin-lattice models 
and performs well for dynamics of both closed and open systems.

\section{Neural Network Quantum States}
\begin{figure*}[t!]
  \includegraphics[width=1.0\linewidth]{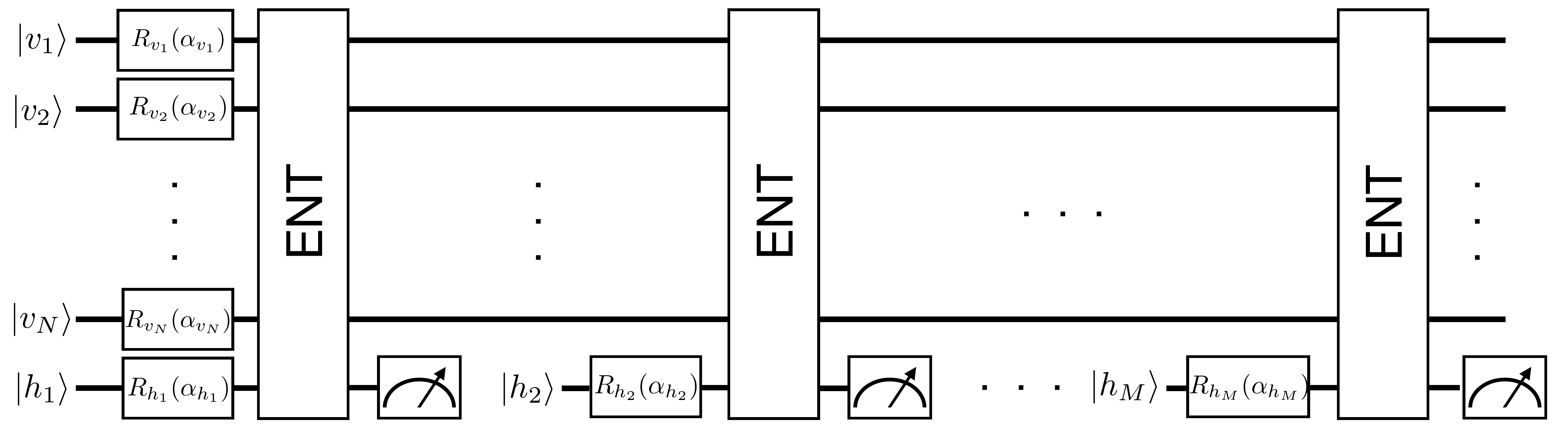}
  \caption{Quantum circuit for preparing uRBM state with qubit recycling scheme described in Eq.~\ref{eq:qrecycle-app}. All the qubits are initialized in $|0\rangle$ state. The single rotations are governed by the relations in Eq.~\ref{eq:q-rotation}. The $j$-th entangling block implements the $\exp(\iu \sum_i  W_{ij}^I \hat v^z_i \hat h^z_j )$ operator  and its explicit form is given in the Appendix \ref{sec:circuits}. After each entangling block, the ancilla qubit representing the $j$-th hidden spin is projected onto $|+\rangle$ state before being recycled.}
  \label{fig:circuit}
\end{figure*}

An RBM quantum state can be obtained from a bipartite Ising Hamiltonian 
\begin{eqnarray}
\hat{H}_{RBM}(\theta) = \sum_{i} b_i \hat{v}^z_i +  \sum_{j} m_j \hat{h}^z_j + \sum_{ij}W_{ij} \hat{v}^z_i \hat{h}^z_j,
\end{eqnarray}
where $\hat v^z_i$ or $\hat h^z_j$ is the Pauli-Z operator for the visible or hidden qubit, respectively. We denote the complex-valued variational RBM parameters as 
$\theta = [b, m, W]$. 
To prepare a complex-valued RBM state using a state preparation protocol proposed in Ref.~\cite{Hsieh2019}, we first entangle $N+M$ qubits (representing $N$ visible and $M$ hidden spins of an RBM architecture) according to

\begin{eqnarray}\label{eq:vhwf}
    |\Psi_{vh}(\theta)  \rangle &=& \frac{e^{\hat{H}_{RBM}(\theta)} }
    {N_{vh}}\ket{++ \dots +}_{vh},
\end{eqnarray}
where $\ket{+}=\frac{1}{\sqrt{2}}(\ket{0}+\ket{1})$, 
$N_{vh} = \sqrt{\prescript{}{vh}{\bra{++ \dots +}} e^{2\hat{H}^R_{RBM}(\theta)}\ket{++\dots+}_{vh}}$, 
$\hat H^R_{RBM}(\theta)$ is the Hermitian part of the RBM Hamiltonian and the subscript $vh$ denotes visible and hidden (ancilla) qubits. 
Eq. \ref{eq:vhwf} gives a conceptually simple wave function that could be generated by first applying single-qubit transformations $\exp(b_i\hat v^z_i )$ and $\exp(m_j\hat h^z_j )$ on individual qubits followed by
$\exp(W_{ij}\hat v^z_i \hat h^z_j)$ to couple the visible and hidden qubits.

Once the extended wave function $\ket{\Psi_{vh}(\theta)}$ is generated, all ancilla qubits (i.e. hidden spins) are post-selected for $\ket{+}_h$ and the desired RBM state is implemented in a quantum circuit
\begin{eqnarray}\label{eq:rbmcircuit}
     |\Psi_v (\theta)  \rangle &=& \frac{\prescript{}{h}{\left\langle ++ \dots + \vert \Psi_{vh}(\theta) \right\rangle}}{N_v},
\end{eqnarray}
where $N_v = \sqrt{ \langle \Psi_{vh}(\theta) | \hat{P}^{(h)}_+  | \Psi_{vh}(\theta)\rangle }$ and $\hat{P}_+^{(h)} = \ket{++\dots+}_h\bra{++\dots+}$ projects all the hidden spins onto $|+\rangle$ state.

Observing that Eq.~\ref{eq:rbmcircuit} can be cast in the following form: 
\begin{eqnarray}\label{eq:qrecycle-app}
\ket{\Psi_v(\theta)}  =  \frac{1}{N_v}
       \left[\prescript{}{}{\bra{+}}\left[e^{\hat h^z_M \left(m_M + \sum_i W_{iM}\hat v^z_i\right)}\right]\ket{+}\right]_{M} \\
       \times \left[\prescript{}{}{\bra{+}}\left[e^{\hat h^z_{M-1}  \left( m_{M-1} + \sum_i W_{iM-1} \hat  v^z_i\right)}\right]\ket{+}\right]_{M-1} \cdots 
       \nonumber \\
        \left[\prescript{}{}{\bra{+}}\left[e^{\hat h^z_1  \left(m_1 + \sum_i W_{i1}\hat v^z_i\right)}\right]\ket{+}\right]_{1}\,\,
       e^{\sum_i b_i \hat v^z_i} \ket{++\cdots+}_v, \nonumber
\end{eqnarray}
where $[ \bra{+}[...] \ket{+}]_j$ encodes the effect of $j$-th hidden spin on all visible spins, 
it is clear that a single ancilla qubit is sufficient to implement the entangling operation sequentially. 

The above quantum operations are non-unitary when RBM parameters are complex, i.e. $b_i^R \neq 0$, $m_j^R \neq 0$ or $W^{R}_{ij} \neq 0$, where we use superscripts $R$ and $I$ to denote the real and imaginary parts of the RBM parameters. 
In particular, the non-unitary two-qubit operations mediating entanglement across the visible-hidden layer are difficult to implement. 
One could adopt a probabilistic scheme~\cite{xia2018rbm} to generate the inter-layer couplings with an extra ancilla qubit. 
However this approach is difficult to scale with the number of qubits since it involves $N*M$ projective measurements. The probability of one successful sampling has therefore a lower bound of $\mbox{e}^{-2\sum_{ij} |w_{ij}|} \sim \mbox{e}^{- O(NM)}$.

For this reason, we only consider the unitary-coupled RBM (uRBM) ansatz in which $W^R_{ij} = 0$ for the rest of this letter~\cite{Hsieh2019}.
Fig.~\ref{fig:circuit} depicts a quantum circuit for preparing a uRBM state composed of $N$ visible spins and $M$ hidden spins.
After initializing all qubits in $\ket{0}$ state, we first perform single qubit rotations representing the terms $\exp(b_i \hat v_i^z)$ and $\exp (m_j \hat h_j^z)$. 
The rotation angles, $\alpha_{v_i / h_j}$, are governed by the relations
\begin{eqnarray} \label{eq:q-rotation}
R_{v_i}(\alpha_{v_i}) |0\rangle &=& \text{e}^{b_i \hat v_i^z}  |+\rangle/c_{v_i}, \\
R_{h_j}(\alpha_{h_j}) |0\rangle &=& \text{e}^{m_j \hat h_j^z}  |+\rangle/c_{h_j}, \nonumber
\end{eqnarray}
where the normalization factors are $c_{v_i}=\sqrt{\bra{+}\exp\left(2 b_i^R \hat v^z_i\right)\ket{+}}$ 
and $c_{h_j}=\sqrt{\bra{+}\exp\left(2 m_j^R \hat h^z_j\right)\ket{+}}$. 
The $j$-th entangling block implements the coupling $\exp(\iu \sum_i W_{ij}^I \hat v^z_i \hat h^z_j )$ and are composed of a series of controlled-Z rotations (see Appendix \ref{sec:circuits} for details). 
Employing the qubit recycling scheme described in Eq.~\ref{eq:qrecycle-app}, 
the ancilla qubit representing the $j$-th hidden spin is projected onto $|+\rangle$ state after each entangling block before being recycled.
Thus we only need $N+1$ qubits in total, and the number of quantum gates is proportional to the number of variational parameters, $N_{var}$, which scales as $O(\alpha N^2)$ where $\alpha = M/N$.

With uRBM, there are only $M$ projective measurements of hidden spins on $\ket{+}$ state, therefore the success probability has improved to $ \mbox{e}^{- O(M)}$. 
We can further mitigate these probabilistic projective measurements. One approach is to re-scale the variational parameters such that the hidden spins remain close to the $\ket{+}$ state~(see Ref. \cite{xia2018rbm}).  
Alternatively, we can use a Monte Carlo scheme with classical post-processing (see Appendix \ref{sec:ensemble}) that enables us to entirely circumvent the post-selection. 

\section{Time-dependent Variational Algorithm}
We adopt a hybrid quantum-classical approach based on the time-dependent variational Monte Carlo (t-VMC) method to simulate the quantum dynamics~\cite{Carleo2012, Carleo2014, Becca2017}.
In the t-VMC framework, we minimize the residue in 
$\text{min}_{\theta}||\iu \frac{\partial | \Psi(\theta) \rangle }{ \partial t}- \hat H_s | \Psi(\theta) \rangle ||,$
where $\hat H_s$ is the system Hamiltonian and the norm is defined as the square root of the inner product. 
The resulting equations of motion for the time-dependent variational parameters are 
\begin{eqnarray}\label{eq:modified_te}
\dot \theta_n =\sum_{m} A_{nm}^{-1} \, \text{Im}[f_m].
\end{eqnarray}
The covariance matrix $A$ and force vector $f$ read
\begin{eqnarray}\label{eq:mtxAvecf}
A_{nm} &=& \text{Re}
\langle \hat O^\dag_n \hat O_m \rangle_v  - \text{Re} \langle \hat O^\dag_n \rangle_v  \text{Re} \langle \hat O_m \rangle_v,\\
f_m &=& 
\langle \hat O^\dag_m \hat H_s \rangle_v - \text{Re} \langle \hat O^\dag_m \rangle_v\langle \hat H_s \rangle_v  , 
\end{eqnarray}
where $\langle \cdots \rangle_v = \bra{\Psi_v(\theta)} \, \cdots \,  \ket{\Psi_v(\theta)}$.  
The derivative operators with respect to the $n$-th variational parameter is defined as $\hat O_n = \frac{ \partial \ln{\ket{\Psi_v(\theta)}}}{\partial \theta_n}$. 
For RBM state defined in Eq.~\ref{eq:rbmcircuit}, the $\hat O_n$ operators can be derived analytically which allows an efficient way of obtaining the gradients
\begin{eqnarray}\label{eq:stochreconfigO}
\hat O_n = \left\{ \begin{array}{ll}
\iu^{1-\delta} \hat v^z_i, & \text{ if } \theta_n = b_i, \\
\iu^{1-\delta} \tanh\left(m_j + \sum_{i} W_{ij} \hat v^z_i\right), & \text{ if } \theta_n = m_j, \\
\iu \hat v^z_i \tanh\left(m_j + \sum_{i}  W_{ij} \hat v^z_i\right), & \text{ if } \theta_n = W_{ij}, \\
\end{array}\right.
\end{eqnarray}
where $\delta = 0$ if $\theta_n = b^{I}_i$ or $\theta_n = m^{I}_j$ or $\theta_n = W^{I}_i$, and $\delta=1$ if $\theta_n = b^{R}_i$ or $\theta_n = m^{R}_j$. 
The variational parameters are updated iteratively according to $\theta_n (t + \delta t) =  \theta_n (t) +  \sum_{m} A_{nm}^{-1} \, \text{Im}[f_m] \delta t$ where $\delta t$ is the update time step.

In conventional t-VMC, the covariance matrix $A$ and force vector $f$ in Eq.~\ref{eq:mtxAvecf} are obtained from Markov chain random walk approach, such as the Metropolis-Hastings algorithm. Such sampling could be challenging for systems that exhibit long correlation time, e.g. systems near phase transition. In a hybrid quantum-classical framework, both $A$ and $f$ are sampled directly from the output of quantum circuit depicted in Fig.~\ref{fig:circuit}, circumventing the difficulties associated with Markov chain methods.

\section{Simulation Results}
To demonstrate the performance of the uRBM algorithm in simulating many-body quantum dynamics, we first consider a 1D transverse-field Ising (TFI) model: 
\begin{eqnarray}
    \hat H_{TFI} &=& - h \sum_{i}\hat \sigma_i^{x} - \sum_{<ij>}\hat \sigma_i^{z}\hat \sigma_j^{z},
\end{eqnarray}
where $h$ denotes the strength of the transverse field. 
Here we study the dynamics of a TFI model induced by quantum quench. 
The TFI system is initially prepared in the ground state for an initial transverse field $h_i$.
The variational parameters of the initial ground state wavefunction are obtained using a hybrid imaginary time algorithm (see Appendix \ref{sec:imaginary_time}). At $t=0$, we introduce an instantaneous change to the transverse field, $h_f \neq h_i$, and let the system evolve under the new Hamiltonian. 

\begin{figure}[t!]
  \centering
  \includegraphics[width=1.00\linewidth]{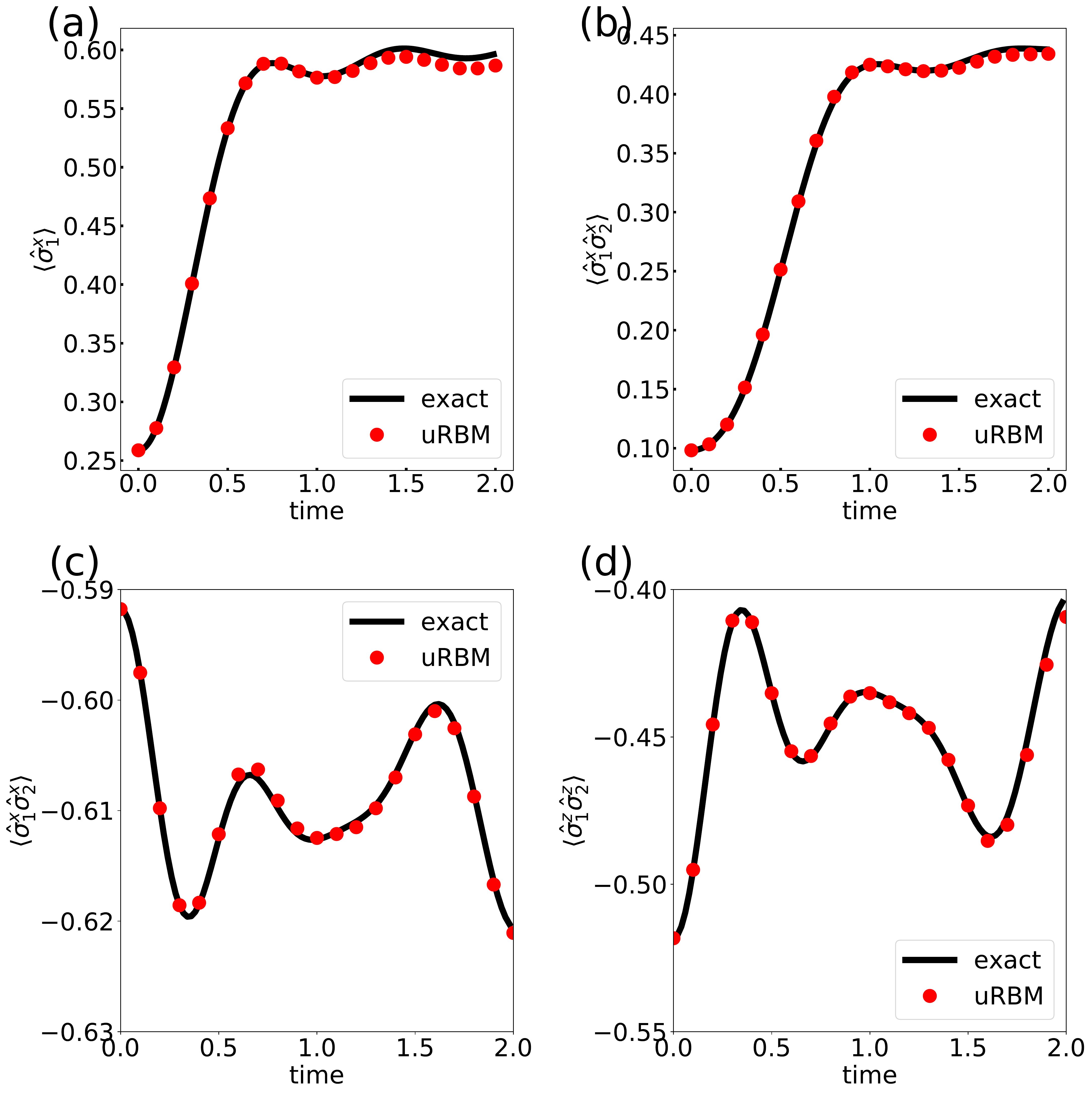}
  \caption{Time evolution induced by quantum quench. Results from uRBM algorithm (symbols) are compared to exact calculations (solids lines). (a) and (b) Dynamics of transverse polarization and its correlation in a 1D Ising model. 
  (c) and (d) Dynamics of magnetization and transverse polarization correlations in a 1D Heisenberg model in a global field.}
  \label{fig:closed_system}
\end{figure}

In Fig.~\ref{fig:closed_system} (a) and (b), we consider a TFI model of 14 spins with periodic boundary condition and the transverse field is changed from $h_i=0.5$ to the critical value of $h_f=1.0$ at $t=0$. In the simulations we use $\delta t = 0.0005$ and $\alpha =M/N = 8$. 
We compare the results from the uRBM algorithm with results from exact diagonalization by studying the evolution of transverse spin polarization $\langle \sigma_1^{x} \rangle$ and its correlation $\langle \sigma_1^{x}  \sigma_2^{x} \rangle$. 
The good agreement with exact results confirms the accuracy of the uRBM algorithm in capturing quantum many-body dynamics. 

Next we consider a 1D anisotropic Heisenberg model with periodic boundary condition in a magnetic field:
\begin{eqnarray}
    \hat H_{H} &=&  - h_z \sum_{i}\hat \sigma_i^{z} + \sum_{<ij>}(J_z \hat \sigma_i^{z}\hat \sigma_j^{z} + \hat \sigma_i^{x}\hat \sigma_j^{x} + \hat \sigma_i^{y}\hat \sigma_j^{y}  ),
\end{eqnarray}
where $h_z$ is the strength of longitudinal field and  $J_z$ is the longitudinal coupling. We perform a quantum quench by instantaneously changing the longitudinal coupling from  $J_z=1.0$ to $J_z=0.5$ at $t=0$. Figs.~\ref{fig:closed_system} (c) and (d) depict the dynamics of spin-spin correlations of a 14-spin Heisenberg model with $h_z=1.0$. 
We use $\delta t =0.0002$ and $\alpha=8$ in our simulations. Again we observe near exact agreement between the results from the uRBM algorithm and exact diagonalization, further confirming the capability of the hybrid uRBM algorithm. 
This can easily be generalized to the more interesting case of random fields in the $z$-direction, which allows the integrable Heisenberg chain to express chaotic behavior and many-body localization \cite{luitz2015many}.
Additionally, we also perform the uRBM simulation of a two dimensional triangular anti-ferromagnetic Ising model (see Appendix \ref{sec:dynamics_triangular}), and again observe excellent agreement with exact calculations.

\section{Open Quantum Systems}
Extending the variational uRBM algorithm to open quantum systems is conceptually straight forward using the stochastic wavefunction approach. 
The dynamics of the density matrix, $\hat \rho$, of an open quantum system can be described by the Linblad master equation~\cite{Breuer2007}
\begin{eqnarray}\label{eq:linblad}
    \frac{d \hat \rho}{dt}  = -\iu[ \hat H_s , \hat \rho] + \frac{1}{2}\sum_k [2\hat L_k \hat \rho \hat L_k^\dagger -  
    \{\hat L_k^\dagger  \hat L_k, \hat \rho \}],
\end{eqnarray}
where $\{.\}$ denotes an anti-commutator, $\hat H_s$ is the system Hamiltonian and the Linblad operators $\hat L_k$ describe the interaction between the system and a Markovian bath.
Instead of solving the Linblad master equation directly, an open quantum system can be equivalently described by an ensemble of pure state trajectories~\cite{Dalibard1992, Carmichael1993}. 
The evolution of these pure state trajectories is governed by a non-Hermitian effective Hamiltonian $\hat H_{eff} = \hat H_s - \frac{\iu}{2}\sum_k (\hat L_k \hat L_k^\dagger - \langle \hat L_k \hat L_k^\dagger \rangle)$ and subject to continuous measurement. The details of implementing these stochastic wavefunction trajectories in quantum circuits can be found in Appendix \ref{sec:stochastic_SE}.

We test the ability of the hybrid uRBM algorithm in simulating the dynamics of an open quantum system by considering a 6-spin 1D TFI model with open boundary condition coupled to a Markovian bath. 
All the spins of the TFI model are initially prepared in $|+\rangle = \frac{1}{\sqrt{2}}(\ket{0} + \ket{1}) $  state. The Linblad operator is $\hat L_k = \sqrt{\gamma} \hat \sigma_k^+$ where $\hat \sigma_k^+$ is a raising operator acting on the $k$-th spin and $\gamma$ determines the strength of system-bath interaction. 
Other parameters used in the simulation are $\alpha= M/N =6$, $\gamma = 0.05$, $h=1.0$, $\delta t =0.0005$.
The dynamics of transverse polarization and its correlation are compared to those from directly solving Eq.~\ref{eq:linblad}. 
It can be seen that the uRBM algorithm is capable of simulating the dynamics of open systems with high accuracy. This further extends the applicability of the hybrid uRBM algorithm to study novel non-equilibrium phenomena in many-body open quantum systems such as phase transitions~\cite{Fink2018, Raftery2014}.

\begin{figure}[t!]
  \centering
  \includegraphics[width=1.0 \linewidth]{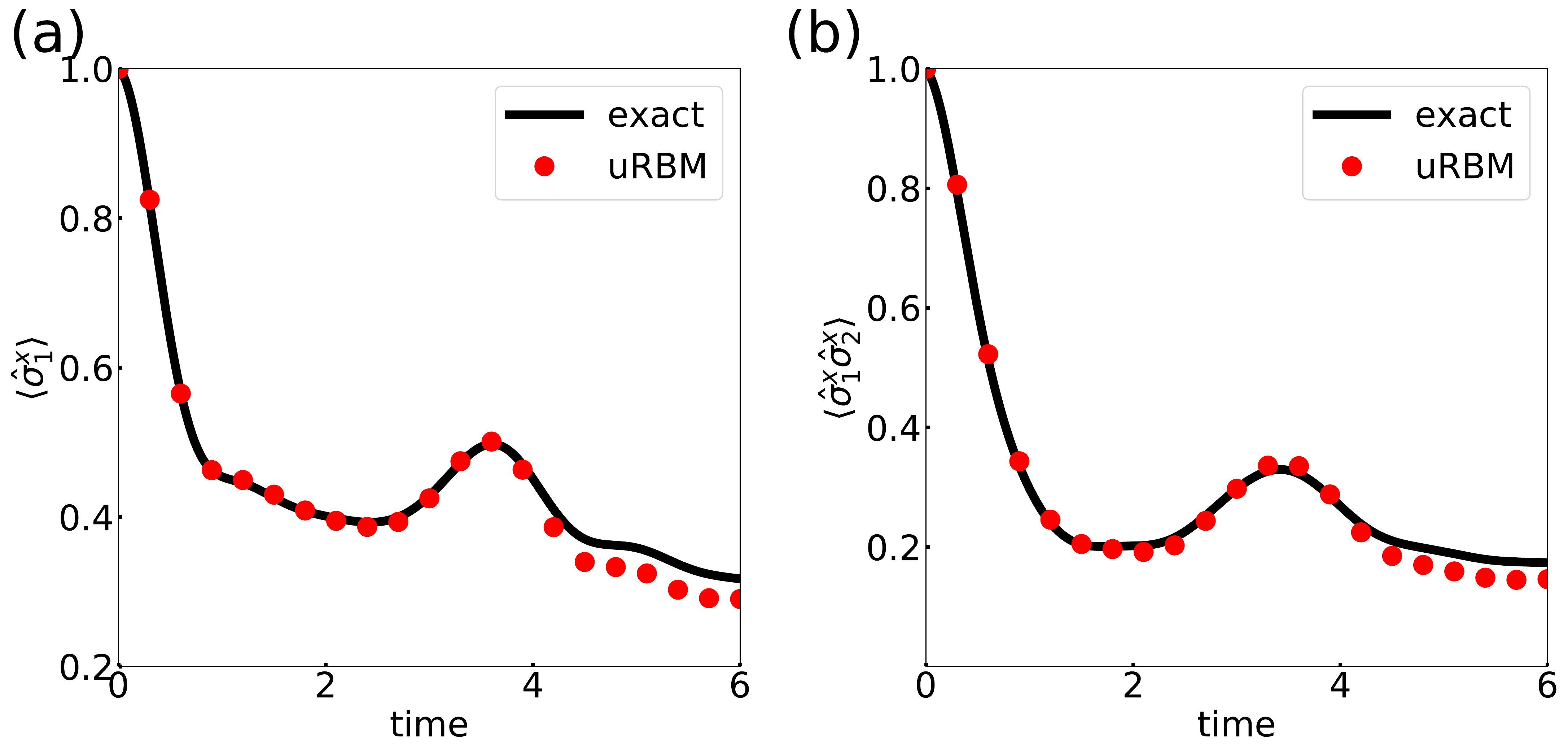}
  \caption{Dynamics of a dissipative 1D Ising model obtained from exact numerical solution of Eq.\ref{eq:linblad} (solid lines) and from the hybrid uRBM algorithm (symbols). The simulation results from the hybrid uRBM algorithm are obtained from averaging over 10000 pure state trajectories, and the error bars are smaller than the symbols}
  \label{fig:dissipation}
\end{figure}

\section{Discussions}
The proposed hybrid uRBM algorithm offers several advantages compared to other NISQ variational algorithms.
First our numerical results up to 18 visible spins (see Appendix \ref{sec:gradient}) show that the gradients in the uRBM ansatz do not decay exponentially with system size, suggesting the absence of the  vanishing gradient (or `barren plateau') issue that affects many variational quantum algorithms~\cite{McClean2018}.
In fact, classical implementations of VMC using RBM ansatz has been demonstrated on systems with more than 100 visible spins~\cite{Zen2020, Zen2020a}.

Second, real and imaginary time variational algorithms \cite{McArdle2019, Endo2020, Yuan2019, Li2017} typically require significantly more measurements (and distinct quantum circuits) than gradient descent approaches such as variational quantum eigensolver (VQE) due of the estimation of covariance matrix $A$. The number of matrix elements in $A$ scales as $O(N_{var}^2)$ where $N_{var}$ is the number of variational parameters.
This measurement cost could be prohibitive in large scale simulation in NISQ devices since $N_{var}$ will be a big number.
Within RBM ansatz all the matrix elements in $A$ can be expressed analytically in terms of the Pauli-Z operators of the visible spins (see Eq.~\ref{eq:stochreconfigO}), a single measurement in the Z-basis contributes to the statistics of all the matrix elements in $A$, thus significantly reducing the number of measurements and distinct circuits required.   

Additionally, the uRBM algorithm offers great flexibility when it comes to the number of ancilla qubits (for hidden spins) and circuit depth. Employing the qubit recycling scheme depicted in Fig.\ref{fig:circuit}, we only need $N+1$ total number of qubits but a circuit depth of $O(\alpha N^2)$ to implement the uRBM state.
At the opposite end of the spectrum, we could use $M$ ancilla qubits to represent $M$ hidden spins, this reduces the circuit depth to $O(N)$, assuming full connectivity like those found in ion-trap based quantum computers~\cite{brown2016iontrap, bruzewicz2019iontrapreview}.
Of course, one could envision an optimal trade-off between qubit number and circuit depth that takes the architecture of the  hardware into account. 
Additionally, we also assess the robustness of our algorithm against imperfections of quantum devices by performing noisy simulations, it is found that the algorithm still yields reliable results in the presence of experimental errors (see Appendix \ref{sec:noisy}).

Finally, the accuracy of the uRBM algorithm can be systematically improved by including more hidden spins. 
For quantum systems that are very strongly correlated, our method can be extended to deep Boltzmann machines (DBM) with modifications. DBM contains more than one layer of hidden spins and has been shown to be able to efficiently represent most quantum states generated by quantum dynamics~\cite{Gao2017, Carleo2018}. The generalization of the variational algorithm to DBM will be presented in a future publication. 

\section{Conclusions}
We have introduced a neural-network based variational quantum algorithm to simulate the dynamics of closed and open quantum many-body systems. 
Our results show that the proposed algorithm is capable of capturing the dynamics of both types of systems with high accuracy.
A key benefit that the integration of quantum devices provides over traditional variational quantum Monte Carlo is the elimination of 
severe ergodicity issues.
Additionally, the proposed variational algorithm offers several advantages over existing NISQ approaches, including absence of barren plateaus for the
considered system sizes, flexibility in qubit-number versus circuit-depth trade-off and low measurement cost. 
These advantages make the algorithm particularly appealing for implementation in NISQ devices.

\textit{Note: }During the preparation of this manuscript, we became aware of related works based on deep quantum feedforward neural networks~\cite{Liu2020} and matrix product states~\cite{Lin2020}.

\section{ACKNOWLEDGMENTS}
We thank L. C. Kwek  for valuable comments and discussions. 

\appendix
\section{Implementation of the entangling gates} \label{sec:circuits}
\begin{figure}[h!]
  \centering
  \includegraphics[width=1.0\linewidth]{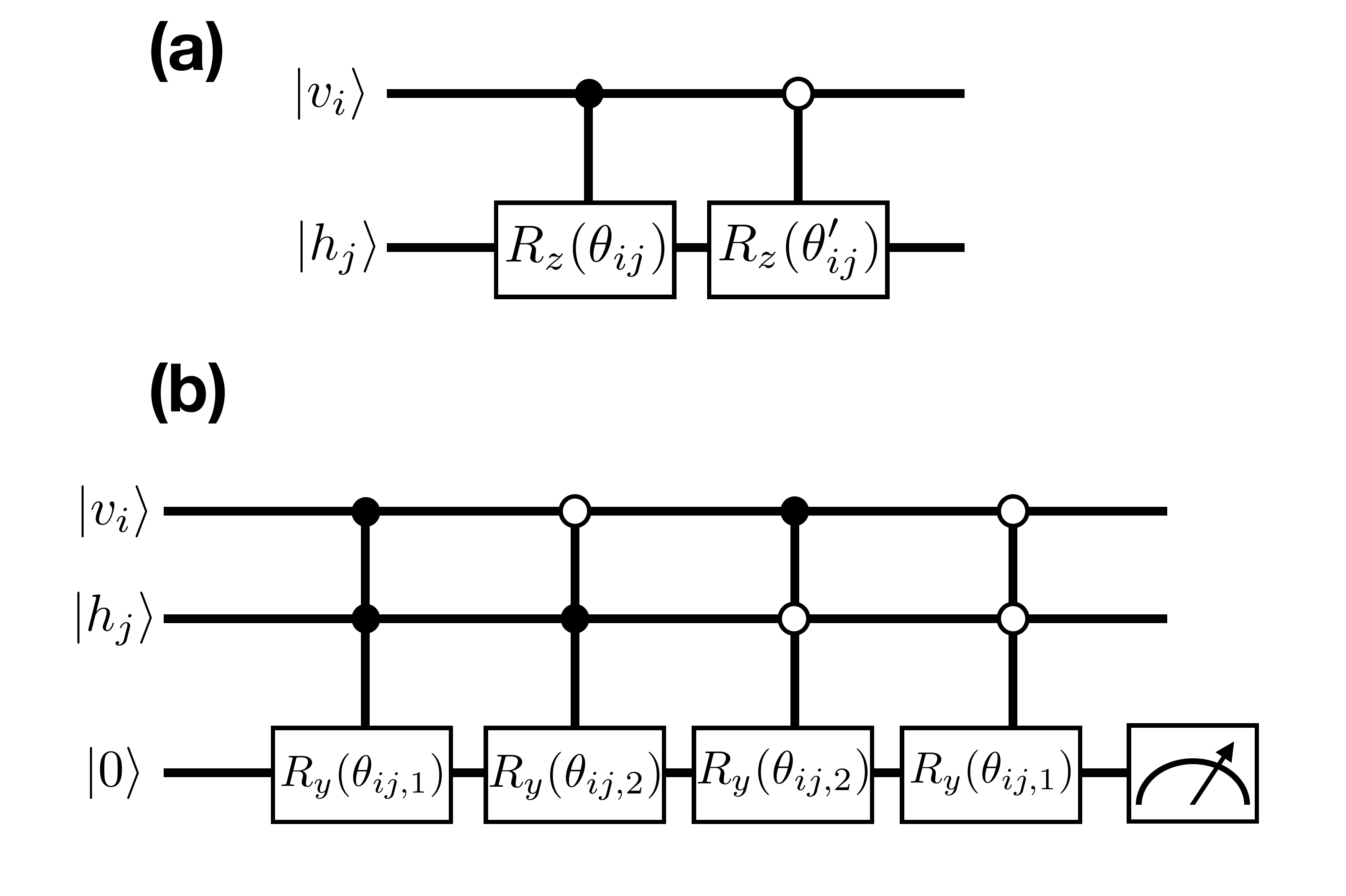}
  \caption{Quantum circuits for the coupling terms (a) $\exp(\iu\sum_{i} W_{ij}^I \hat v^z_i \hat h^z_j )$ and (b) $\exp(\sum_{i} W_{ij}^R \hat v^z_i \hat h^z_j )$  between $i$-th visible and $j$-th hidden spins, .  }
  \label{fig:ent}
\end{figure}

For unitary coupled RBM (uRBM) ansatz ($W_{ij}^R = 0$), the $j$-th entangling block in the quantum circuit of Fig.~1 in the main text implements the operation $\exp(\iu\sum_{i} W_{ij}^I \hat v^z_i \hat h^z_j )$ that couples  the $j$-th hidden spin with all the visible spins. The quantum circuit for each coupling term $\exp(\iu W_{ij}^I \hat v^z_i \hat h^z_j )$ is shown in Fig.\ref{fig:ent}(a) where $\theta_{ij} = - \theta'_{ij} = -W_{ij}^I$. 
For full RBM states with complex value couplings,  the non-unitary operation $\exp(W_{ij}^R \hat v^z_i \hat h^z_j )$ can be implemented using the probabilistic scheme introduced by Xia and Sabre \cite{xia2018rbm} to generate the inter-layer couplings with an extra ancilla qubit. The quantum circuit of this scheme is shown in Fig.\ref{fig:ent}(b). The rotation angles in the controlled gates are 
\begin{eqnarray}
\theta_{ij, 1} = 2\sin^{-1} (\sqrt{\exp(W^R_{ij} - |W^R_{ij}|)}), \\ 
\theta_{ij, 2} = 2\sin^{-1} (\sqrt{\exp(-W^R_{ij} - |W^R_{ij}|)}). \nonumber
\end{eqnarray}
After each operation $\exp(\sum_i W_{ij}^R \hat v^z_i \hat h^z_j )$ is implemented, the ancilla qubit is measured. If the ancilla qubit is in state $|1 \rangle$, we continue to the next coupling term, otherwise we start from the beginning.
Given the $N*M$ number of probabilistic measurements of the ancilla qubit, this approach is difficult to scale with the number of qubits for large scale simulation. 

\section{Ensemble Preparation of unitary RBM States} \label{sec:ensemble}
Here we discuss an ensemble preparation of the unitary RBM state without resorting to the probabilistic post-selection of hidden spins~\cite{Hsieh2019}. First we note that each term on the right hand side of Eq.~4 in the main text can be written as 
\begin{eqnarray}\label{eq:block-decomp}
 &&\prescript{}{}{\bra{+}}\left[e^{\hat h^z_j \left(m_j + \sum_i iW^{\text{I}}_{ij}\hat v^z_i\right)}\right]\ket{+} \\
& = & \sum_{s=\pm} \bra{+} e^{m^{\text{R}}_j \hat h^z_j} \ket{s} \bra{s} e^{\left(im^{\text{I}}_j+\sum_i iW^{\text{I}}_{ij}\hat v^z_i\right) \hat h^z_j} \ket{+} \nonumber \\
& = & \sum_{s=\pm} R_{s}(m_j^{\text{R}}) \bra{s} e^{\left(im^{\text{I}}_j+\sum_i iW^{\text{I}}_{ij}\hat v^z_i\right) \hat h^z_j} \ket{+}, \nonumber
\end{eqnarray}
where $R_{s}(m^{\text{R}}_j)=\bra{+} e^{m^{\text{R}}_j \hat h^z_j} \ket{s} $ can be computed classically as it only involves single qubit operation. Using Eq.~\ref{eq:block-decomp}, we re-write Eq.~4 in the main text such that
\begin{eqnarray}\label{eq:qrecycle2}
\ket{\Psi_v(\theta)} & = &  \sum_{s_M=\pm} \cdots \sum_{s_1=\pm}  \frac{1}{N_v}\left(\prod_{j=1}^MR_{s_j}(m^{\text{R}}_j)\right) \\
       &&\prescript{}{}{\bra{s_M}}\left[e^{\hat h^z_M \left(i m^{\text{I}}_M + \sum_i iW^{\text{I}}_{iM}\hat v^z_i\right)}\right]\ket{+} \cdots \nonumber \\
       \nonumber \\
       & &  \prescript{}{}{\bra{s_1}}\left[e^{\hat h^z_1  \left(i m^{\text{I}}_1 + \sum_i iW^{\text{I}}_{i1}\hat v^z_i\right)}\right]\ket{+} \,\,
       e^{\sum_i b_i \hat  v^z_i} \ket{++\cdots}_v \nonumber \\
& = & \sum_{s_M=\pm} \cdots \sum_{s_1=\pm} \frac{N_{\vec s}}{N_v}\left(\prod_{j=1}^MR_{s_j}(m^{\text{R}}_j)\right )\ket{\Psi_v^{\vec{s}}(\theta)}, \nonumber
\end{eqnarray}
where $\vec{s}=[s_1,\cdots,s_M]$ and $N_{\vec s}$ is the normalization to ensure
$\bk{\Psi_{v}^{\vec{s}}}{\Psi_{v}^{\vec{s}}}=1$. $\ket{\Psi_v^{\vec{s}}(\theta)}$ is a visible-spin wave function created by projecting hidden spins onto basis states $\ket{s_1\cdots s_M}_h$. Therefore the state preparation protocol given in Eq.~\ref{eq:qrecycle2} replaces the  probabilistic post-selection of hidden spins with a summation over all possible $\vec{s}$ of hidden spins.

The expectation value of an observable $\hat O$ can be calculated 
\begin{eqnarray}\label{eq:expo}
&&\bra{\Psi_v(\theta)} \hat O \ket{\Psi_v(\theta)}   \\
 &&=\int dz \vert\bk{\mathbf z}{\Psi_v(\theta)}\vert^2  \left[\int dz^\prime  O(z,z^\prime) \frac{\bk{\mathbf z^\prime}{\Psi_v(\theta)}}{\bk{\mathbf z}{\Psi_v(\theta)}}\right].  \nonumber
\end{eqnarray}
The above equation suggests that the expectation value of an observable $\hat O$ can be turned into the average of the expression inside the square bracket if we can efficiently sample $z$ according to the probability density $|\bk{\mathbf z}{\Psi_v(\theta)}|^2$.

\section{Variational imaginary time evolution} \label{sec:imaginary_time}
\begin{figure}[h!]
  \centering
  \includegraphics[width=1.0\linewidth]{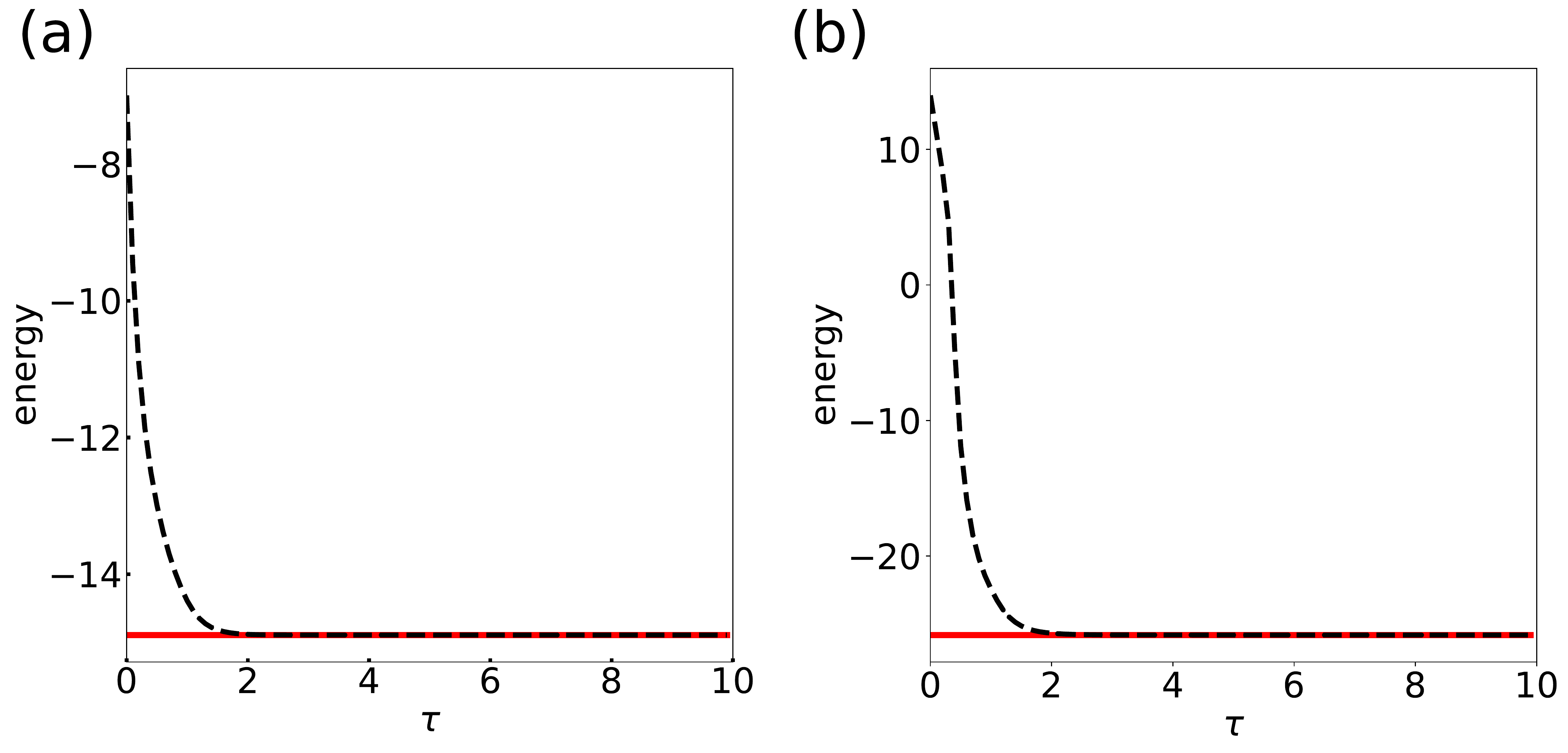}
  \caption{Imaginary time evolution of (a) 1D Ising model and (b) Heisenberg models.
  The solid lines are the exact ground-state energy. The dashed black lines represent the imaginary time evolution using the variational uRBM algorithm.}
  \label{fig:ITE}
\end{figure}

In the numerical examples of closed systems in Fig. 2 of the main text, the initial states are prepared as the ground states of the initial Hamiltonians before quantum quenches.
The variational parameters of these initial wavefunctions are obtained via a variational quantum-classical imaginary time evolution (ITE) following the Stochastic Reconfiguration framework~\cite{Sorella2000}.
The update rule of the variational parameters in the hybrid ITE algorithm is 
\begin{eqnarray}\label{eq:modified_ite}
\dot \theta_n (\tau)=\sum_{m} A_{nm}^{-1} \, \text{Re}[f_m].
\end{eqnarray}
where $\tau$ denotes the imaginary time, the definitions of the covariance matrix $A$ and the force vector $f$ are the same as the real time algorithm (i.e.  Eqs. (7) and (8)) in the main text. The parameters are updated iteratively 
\begin{eqnarray}\label{eq:ite_update}
\theta_n (\tau + \delta \tau) =  \theta_n (\tau) + \delta \tau A^{-1} \text{Re}[f]
\end{eqnarray}
where $\delta \tau$ is the imaginary time step. In our simulations, we use $\delta \tau = 0.01$ for 2500 steps. 
At $\tau =0$, the variational RBM parameters are initialized as Gaussian random numbers with zero mean and variance of 0.01. 
The imaginary time evolution of the 14-spin 1D Ising and Heisenberg models used in the main text are shown in Fig.~\ref{fig:ITE}. 

\section{Dynamics of Triangular Anti-Ferromagnetic Lattice} \label{sec:dynamics_triangular}
\begin{figure}[h!]
  \centering
  \includegraphics[width=1.0\linewidth]{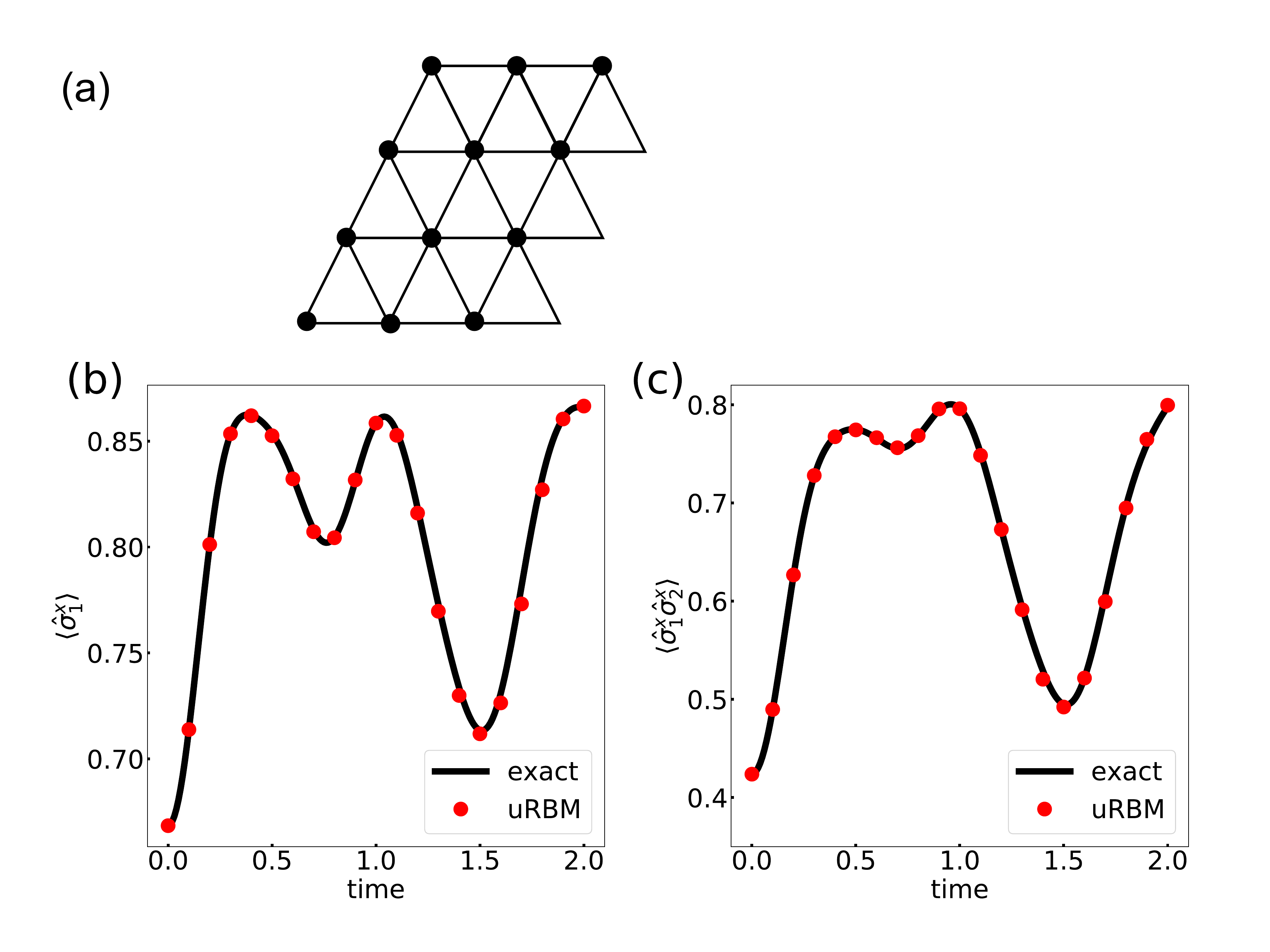}
  \caption{Time evolution in a two dimensional triangular anti-ferromagnetic Ising lattice induced by quantum quench. (a) Configuration of the triangular lattice with periodic boundary condition. (b) and (c) Results from uRBM algorithm (symbols) are compared to exact calculations (solids lines) for dynamics of transverse polarization and its correlation.}
  \label{fig:triangular}
\end{figure}

Here we consider the dynamics of a two dimensional triangular anti-ferromagnetic Ising (TAFI) model  with periodic boundary condition, a system known for critical slow down for a range of magnetic fields. The Hamiltonian is given by 
\begin{eqnarray}
    \hat H &=& - h \sum_{i}\hat \sigma_i^{x} + \sum_{<ij>}\hat \sigma_i^{z}\hat \sigma_j^{z},
\end{eqnarray}
where $h$ denotes the strength of the transverse field. 
A schematic of the triangular lattice is shown in Fig.~\ref{fig:triangular} (a). 
 We perform a quantum quench by instantaneously changing the transverse field from  $h=0.5$ to $h=1.0$ at $t=0$. We use $\delta t =0.0005$ and $\alpha= M/N = 8$ in our simulations. Figs.~\ref{fig:triangular} (b) and (c) shows the dynamics of transverse spin polarization $\langle \sigma_1^{x} \rangle$ and its correlation $\langle \sigma_1^{x}  \sigma_2^{x} \rangle$ of a 12-spin triangular lattice.  
The good agreement with exact results for this more challenging example further demonstrate the capability of the uRBM algorithm in capturing quantum many-body dynamics. 

\section{Ergodicity Problem in Triangular Anti-Ferromagnetic Lattice}
\begin{figure}
    \centering
    \includegraphics[width=.7\linewidth]{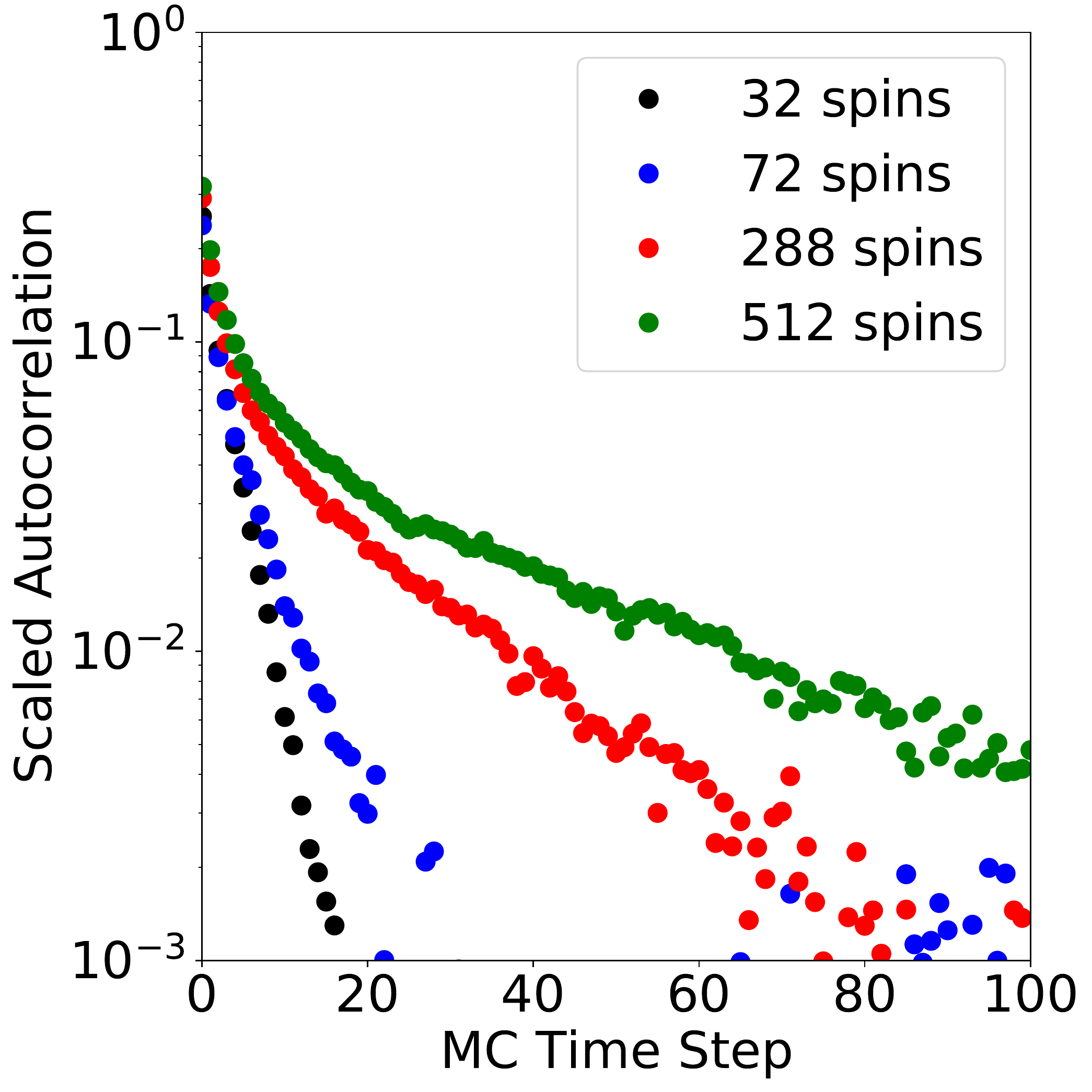}
    \caption{Autocorrelation of a two dimensional triangular anti-ferromagnetic Ising model as a function of Monte Carlo step number for
    $O=\sigma^z_{(0,0)}\sigma^z_{(L/2,L/2)}$. $y$-aixs is scaled by $\langle O(\tau=0)O(\tau=0)\rangle$
    to ensure that maximum value is unity.}
    \label{fig:TAFI}
\end{figure}

Here we investigate the ergodicity issue of TAFI model in the classical limit (i.e. $h=0$).  
The classical TAFI model is one of the
simplest examples of a frustrated magnet hosting a spin liquid phase at zero temperature
\cite{blote1982roughening}. The large correlation lengths associated with scale
invariant behavior close to such phases lead to complex energy landscapes and a poor
performance of simple Metropolis like updates in Monte Carlo. Although it is possible
in special cases to develop efficient cluster algorithms, most frustrated spin systems
do not lend themselves to such methods. This is made explicit for the TAFI in a uniform
transverse field in Ref. \cite{biswas2016quantum},
where the authors develop a specialized cluster
algorithm to study the physics at low transverse fields. To quantify the performance
of standard Metropolis updates on the classical TAFI, we calculate an autocorrelation
function of the spin correlation on lattice sites with maximal separation,
i.e. $\sigma^z_{(0,0)}\sigma^z_{(L/2,L/2)}$, where the subscripts denote the spin position and $L$ is the lattice length in each dimension. This is shown for a range of sizes in
Fig.~\ref{fig:TAFI} and we see that the time to equilibrium grows with system size.
As qualitative features of the ground state phase remain similar at finite
transverse fields, we expect that similarly long autocorrelation times would be
seen in that case as well, evidence for the same is shown explicitly in Ref.~\cite{biswas2016quantum}.
With direct sampling in quantum computers, we would circumvent this ergodicity issue. 

\section{Numerical Simulations with Gaussian Noise} \label{sec:noisy}
\begin{figure}[th!]
  \centering
  \includegraphics[width=1.0\linewidth]{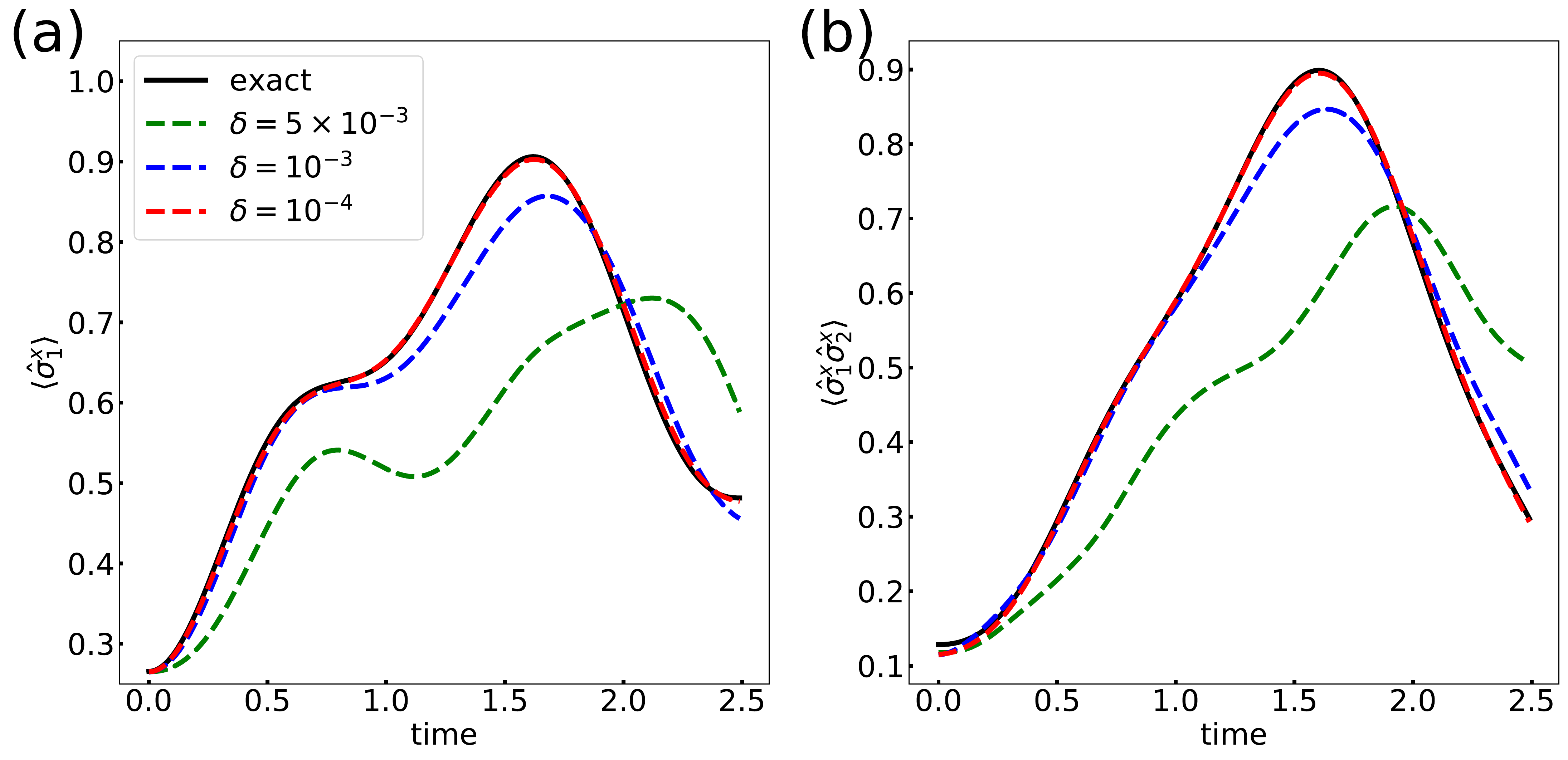}
  \caption{Noisy simulations of a 6-spin 1D Ising  lattice induced by quantum quench. 
  To account for the imperfections of actual quantum devices, random Gaussian noises with standard 
deviation $\delta$ are added to the covariance matrix, $A$ and the force vector $f$. 
 Results from noisy uRBM simulations (dahsed lines) are compared to exact calculations (black solid lines) for dynamics of transverse polarization and its correlation.}
  \label{fig:noisy}
\end{figure}
To assess the robustness of the neural-network variational algorithm against imperfections of near-term quantum computers and errors due to finite number of measurements, we perform noisy simulations by adding random Gaussian noise into the matrix elements of the covariance matrix, $A$ and the force vector $f$ at each time step $\delta t$.
We use a Gaussian random number of zero mean and standard deviation of $\delta$.  
 We perform noisy simulation of a 6-spin 1D Ising model (see Eq.~10 of main text for Hamiltonian)and study its dynamics upon quantum quench when the transverse field is changed from $h_i=0.5$ to $h_f=1.0$ at $t=0$, the time evolution of transverse polarization and its correlation is shown in Fig.~\ref{fig:noisy}. 
It can be seen that our algorithm is robust against small errors, but as the magnitude of the noise increases, the quantum dynamics start to deviate significantly from the exact dynamics. 

As the performance of quantum computers improves rapidly in recent years, error rates of $10^{-4} - 10^{-3}$ used in Fig.~\ref{fig:noisy} can be expected the near-future. Particularly single-qubit gate fidelity of $99.9999\%$~\cite{Harty2014} and two-qubit gate fidelity of $99.9\%$~\cite{Ballance2016, Gaebler2016} have already been demonstrated
in trapped ion quantum computer.


 
\section{Derivations of wavefunction derivatives} \label{sec:derivatives}
The derivative of $\ket{\Psi_v(\theta)}$ can be written as 
\begin{eqnarray}\label{eq:chainrule2}
    \left \vert \frac{\partial \Psi_v}{\partial \theta_n} \right \rangle &=& \frac{_h\langle ++\dots + |\partial_{\theta_n} \tilde \Psi_{vh}\rangle}{\tilde N_v} 
     \\
     &-& Re \left( \frac{ \langle \tilde \Psi_{vh}|}{ \tilde N_v} \hat{P}_+^{(h)} \frac{| \partial_{\theta_n} \tilde \Psi_{vh}  \rangle}{\tilde N_v} \right)   \frac{_h\langle ++\dots + | { \tilde \Psi_{vh}}  \rangle}{ \tilde N_v}, \nonumber
\end{eqnarray}
where $\ket{\tilde\Psi_{vh}(\theta)} =  e^{\hat H_{RBM}(\theta)} \ket{++\dots+}_{vh}$ is the unnormalized wavefunction and $ N_v = \sqrt{\bra{\Psi_{vh}(\theta)}P^{(h)}_+\ket{\Psi_{vh}(\theta)}}$. 
$\hat H_{RBM}(\theta,h)$ is the RBM Hamiltonian with the hidden spins $\hat h^z_j$ replaced with binary values of $\pm 1$.
The derivatives of $\ket{\tilde \Psi_{vh}(\theta)}$ are in turn given by
\begin{eqnarray}\label{eq:unvhwf-deriv}
\frac{ \partial |\tilde\Psi_{vh} \rangle }{\partial b^{R}_i} &=&  \hat{v}^z_i   |\tilde\Psi_{vh} \rangle , \\
\frac{ \partial |\tilde\Psi_{vh} \rangle }{\partial m^{R}_j} &=& \tanh\left(m_j + \sum_i W_{ij} \hat{v}^z_i\right)  |\tilde\Psi_{vh} \rangle, \nonumber \\
\frac{ \partial |\tilde\Psi_{vh} \rangle }{\partial W^{R}_{ij}} &=& \hat{v}^z_i \tanh\left(m_j + \sum_i W_{ij} \hat{v}^z_i\right)   |\tilde\Psi_{vh} \rangle, \nonumber \\
\frac{ \partial |\tilde\Psi_{vh} \rangle }{\partial b^{I}_i} &=&  i\hat{v}^z_i | \tilde \Psi_{vh} \rangle, \nonumber \\
\frac{ \partial |\tilde\Psi_{vh} \rangle }{\partial m^{I}_j} &=&  i \tanh\left(m_j + \sum_i W_{ij} \hat{v}^z_i\right)   |\tilde\Psi_{vh} \rangle, \nonumber \\
\frac{ \partial |\tilde\Psi_{vh} \rangle }{\partial W^{I}_{ij}} &=&  i  \hat{v}^z_i \tanh\left(m_j + \sum_i W_{ij} \hat{v}^z_i\right) |\tilde\Psi_{vh} \rangle. \nonumber
\end{eqnarray}
Substituting Eqs.~\ref{eq:chainrule2}-\ref{eq:unvhwf-deriv} into in the derivative operator, 
$O_n = \frac{\partial \ln\ket{ \Psi_v}}{\partial \theta_n}$, 
we arrive at Eq.(9) in the main text. 

\section{Measuring Derivatives in Quantum Circuits} \label{sec:measure_derivatives}
Here we explain how to measure the matrix elements of the covariance matrix $A$ and force vector $f$. Since $[\hat O_n,\hat{v}^z_i]=0$, the expectation values of $\langle \hat O^\dag_n \hat O_m\rangle_v$ can be obtained by measuring the visible spins in the $z$-basis
\begin{eqnarray}
&&\langle \hat O^\dag_n \hat O_m\rangle_v \\
& = & \bra{\Psi_v(\theta)} \hat O^\dag_n \hat O_m  \ket{\Psi_v(\theta)} \nonumber \\
& = &  \sum_{\mathbf{z}_v} \vert \bra{\Psi_v(\theta)}  \mathbf{z}_v \rangle \vert^2 \hat O^\dag_n(\mathbf{z}_v) \hat O_m(\mathbf{z}_v)  \nonumber \\
&& \xrightarrow[\text{according to } P_v(\mathbf{z}_v)]{\text{Monte Carlo sampling}}  \sum_{k=1}^{N_{\text{exp}}} \frac{\hat O^\dag_n(\mathbf{z}_v^k) \hat O_m(\mathbf{z}_v^k) }{N_{\text{exp}}}, \nonumber
\end{eqnarray}
where we have inserted the completeness relation, $\sum_{\mathbf{z}_v} \ket{{\mathbf{z}_v}} \bra{{\mathbf{z}_v}}$, into the second line.
Here $P_v(\mathbf{z}_v) = \vert \bk{\Psi_v(\theta)}{\mathbf{z}_v} \vert^2$ is a probability density, $\mathbf{z}_v = [z_{v,1}, \cdots, z_{v,N}]$ is a 
length-$N$ binary string, and $\hat O_{n}$ is defined in Eq.~9 of the main text, with the visible spin operators replaced by $\mathbf{z}_v$. 
$N_{\text{exp}}$ samples of [$\mathbf{z}_v^{(k=1)} \cdots  \mathbf{z}_v^{(k=N_\text{exp})}$] are obtained from the quantum circuit to estimate $\hat O^\dag_n (\mathbf{z}_v^k) \hat O_m(\mathbf{z}_v^k)$, according to the third line (the Monte Carlo method) in equation above.  The expression $\hat O^\dag_n (\mathbf{z}_v^k) \hat O_m(\mathbf{z}_v^k)$ can be evaluated efficiently once a computational state $\mathbf{z}^k_v$ is specified.
$\langle \hat O_n \rangle_v$ can be similarly calculated. 

The evaluation of $\langle \hat O^\dag_m \hat H \rangle_v$ is more complicated,
\begin{eqnarray}\label{eq:mc2}
&&\langle \hat O^\dag_m \hat H \rangle_v \\
& = &  \bra{\Psi_v(\theta)}  \hat O^\dag_m \hat H  \ket{\Psi_v(\theta)} \nonumber \\
& = &   \sum_{\mathbf{z}_v,\tilde{\mathbf{z}}_v} \bk{\Psi_v(\theta)}{\mathbf{z}_v}  O^\dag_m(\mathbf{z}_v) \hat H(\mathbf{z}_v,\tilde{\mathbf{z}}_v)
\bk{\tilde{\mathbf{z}}_v}{\Psi_v(\theta)} \nonumber \\
& = & \sum_{\mathbf{z}_v} \vert \bk{\Psi_v(\theta)}{\mathbf{z}_v} \vert^2 \left( \sum_{\tilde{\mathbf{z}}_v}   \hat O^\dag_m(\mathbf{z}_v) \hat H(\mathbf{z}_v,\tilde{\mathbf{z}}_v)
\frac{\bk{\tilde{\mathbf{z}}_v}{\Psi_v(\theta)}}{\bk{\mathbf{z}_v}{\Psi_v(\theta)}}\right), \nonumber \\
&& \xrightarrow[\text{according to } P_v(\mathbf{z}_v)]{\text{Monte Carlo sampling}}  \sum_{k=1}^{N_{\text{exp}}} \frac{1}{N_{\text{exp}}} \times \nonumber\\
&&\,\,\,\,\,\,\,\,\,\,\,\,\,\,\,\,\,\,\,\,\,\,\,\,\,\,\,\,\,\,\,\,\,\,\, \left( \sum_{j}   \hat O^\dag_m(\mathbf{z}^k_v) \hat H(\mathbf{z}^k_v,\tilde{\mathbf{z}}^{k,j}_v)\frac{\bk{\tilde{\mathbf{z}}^{k,j}_v}{\Psi_v(\theta)}}{\bk{\mathbf{z}^k_v}{\Psi_v(\theta)}}\right), \nonumber
\end{eqnarray}
where $\hat H(\mathbf{z}_v^k, \tilde {\mathbf{z}}^{k,j}_v) = \langle \mathbf{z}^k_v \vert \hat H \vert \tilde{\mathbf{z}}^{k,j}_v \rangle$. For physical systems, the Hamiltonian, $ \hat H=\sum_l w_l 
\hat P_l$, is  a linear combination of Pauli strings, i.e. $\hat H$ is a sparse matrix such that each computational state $\ket{\mathbf{z}^k_v}$ is only connected to 
a few other states $\ket{ \tilde{\mathbf{z}}_v^{k,j}}$. 
The expression in the bracket of the third line of Eq.~\ref{eq:mc2} can then be evaluated classically efficiently. 

\section{Gradients of uRBM Parameters} \label{sec:gradient}
\begin{figure}[h!]
  \centering
  \includegraphics[width=1.0\linewidth]{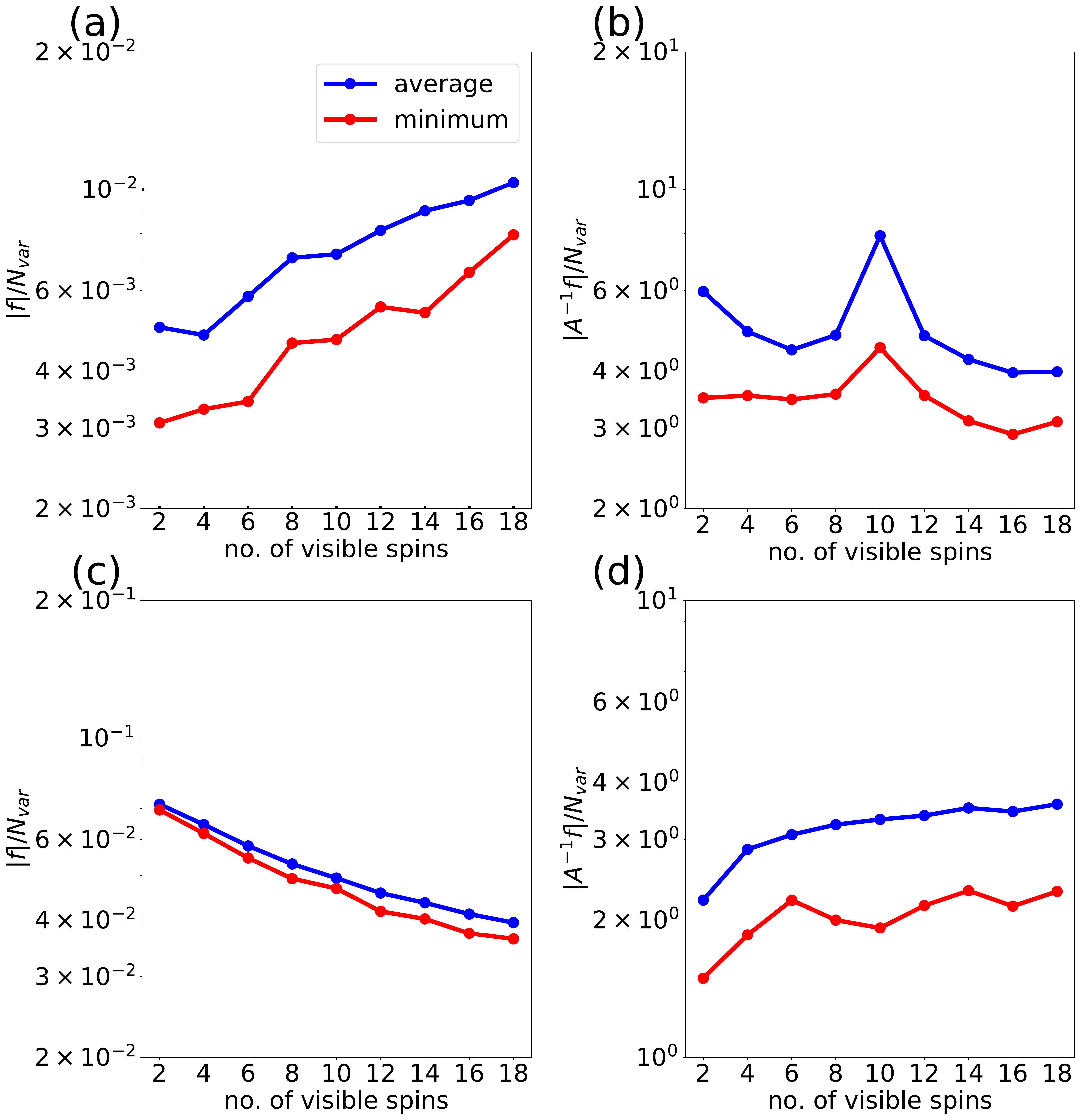}
  \caption{The norms of $f$ and $A^{-1}f$ as a function of system size for a 1D transverse field Ising model ((a) and (b)) and a 1D Heisenberg model in longitudinal field ((c) and (d)). The norms are normalized by the number of variational parameters. The blue and red lines denote the average and minimum from 100 random initializations, respectively.}
  \label{fig:gradient}
\end{figure}

Here we show that the proposed uRBM algorithm does not suffer from the `barren plateau' issue that affects many variational quantum algorithms~\cite{McClean2018}. 
In Fig.~\ref{fig:gradient} we plot the norms of the force vector, $f$, and the gradient, $A^{-1}f$, as a function of system size, both quantities are normalized by the total number of variational parameters. 
We consider a 1D transverse field Ising model (panels (a) and (b)) and a 1D Heisenberg model in longitudinal field (panels (c) and (d)) with periodic boundary condition, both with magnetic field strength $h=1.0$.
The number of hidden spins is fixed at $M=6$. The RBM parameters are randomly initialized as Gaussian variables with variance of 0.01, the blue and red lines in Fig.~\ref{fig:gradient} denote the average and minimum of 100 random initializations, respectively.

The force vector, $f$, is simply the gradient vector of the energy function, $E_\theta = \bra{\Psi_v}(\theta) H_s \ket{\Psi_v}(\theta)$, whereas the real and imaginary parts of $A^{-1}f$ dictate the parameter update in imaginary and real time evolution (see Eq.~6 in main text and Eq. \ref{eq:modified_ite} above), respectively. From Fig.~\ref{fig:gradient}, we can clearly see that both $f$ and $A^{-1}f$ do not decay exponentially with system size, indicating the uRBM algorithm does not suffer from the vanishing gradient ( or `barren plateau') issue.

\section{Stochastic Schr\"{o}dinger Equation} \label{sec:stochastic_SE}
The dynamics of an open quantum system coupled to a Markovian bath can be described by an ensemble of pure state trajectories under continuous measurement~\cite{Dalibard1992, Carmichael1993}. 
The stochastic differential equation governing the evolution of the pure state trajectory can be written as 
\begin{eqnarray}\label{eq:stochastic_se}
   d \ket{\psi (t)} &=& -\iu \hat H_{eff} \ket{\psi(t) } dt +  \\
               &&  \sum_k \Big( \frac{ \hat L_k \ket{\psi(t)}}{\|\hat L_k \ket{\psi(t)}\|} 
                - \ket{\psi(t)} \Big) dN_k(t), \nonumber
\end{eqnarray}\label{eq:effective_H}
where the non-Hermitian effective Hamiltonian
\begin{eqnarray} \label{eq:effective_H}
\hat H_{eff} = \hat H_s - \frac{\iu}{2}\sum_k (\hat L_k \hat L_k^\dagger - \langle \hat L_k \hat L_k^\dagger \rangle),
\end{eqnarray}
describes the deterministic evolution of the trajectory.
The first term on the right hand side of Eq.~\ref{eq:effective_H} is the usual system Hamiltonian, and the non-Hermitian part (terms in bracket) describes the damping process. 
The terms $\langle \hat L_k \hat L_k^\dagger \rangle = \langle\psi(t)| \hat L_k \hat L_k^\dagger|\psi(t) \rangle$ ensure normalization of the wavefunction. 
The deterministic evolution is interrupted by instantaneous changes to the wavefunction, $\ket{\psi} \rightarrow \frac{\hat L_k \ket{\psi}}{\|\hat L_k \ket{\psi}\|}$, the so-called quantum jumps described by the second term on the right hand side of Eq.~\ref{eq:stochastic_se}.
The random numbers $dN_k(t)$ associated to the jumps take on the values of $0$ or $1$ and have expectation values of 
\begin{eqnarray}\label{eq:jump_prob}
   E[dN_k(t)] = \bra{\psi(t)}\hat L_k^\dagger \hat L_k \ket{\psi(t)} dt.
\end{eqnarray}
$E[dN_k(t)]$ represents the probability of a quantum jump associated to the Linblad operator $\hat L_k$, the total jump probability is thus given by $\sum_k E[dN_k(t)]$.

Next we describe how the stochastic Schr\"{o}dinger equation can be simulated using variational algorithm described in the main text. 
We first assume that the wavefunction at time $t$, $\ket{\psi(t)}$, can be represented by a parametrized ansatz $\ket{\Psi(\theta)}$ prepared in a quantum circuit. 
Between quantum jumps, the deterministic part of the stochastic Schr\"{o}dinger can then be simulated with Eqs.~6-8 in the main text, but replacing the system Hamiltonian with the effective Hamiltonian $H_{eff}$. 
To realize a quantum jump associated with $\hat L_k = \sigma_k^+$, we first note that the raising operator can be written as $\hat \sigma^+_k = \text{e}^{- \tau \hat H_k } \text{e}^{- \iu \frac{\pi}{2} \hat \sigma^x_k}$ for large enough $\tau$~\cite{Endo2020} and $\hat H_k = \ket{0}_k\bra{0}$.
Then the quantum jump can be realized in a quantum circuit by evolving the quantum state s $\sigma^x_k$ for duration $\frac{\pi}{2}$. 
Then we propagate the state by imaginary time evolution under $\hat H_k$ for $\tau$. 
In our simulations we use $\tau =20$ and time step of $\delta \tau =0.01$. 

 \bibliography{qrbm}
\end{document}